\documentclass[twocolumn,showpacs,preprintnumbers,amsmath,amssymb,10pt]{revtex4}
\usepackage{amsmath,epsfig,graphicx,amssymb}

\begin{document}

\title{\Large \bf The Adiabatic Transport of Bose-Einstein Condensates
in a Double-Well Trap: Case a Small Nonlinearity}

\author{\bf  V.O. Nesterenko $^1$, A.N. Novikov $^1$, and E. Suraud $^2$}

\date{\today}

\address{\it$^{1}$ Laboratory of Theoretical Physics, Joint Institute
for Nuclear Research, Dubna, Moscow region, 141980, Russia;
e-mail: nester@theor.jinr.ru, novikov@theor.jinr.ru}
\address{\it$^{2}$ Laboratoire de Physique
Thoretique, Universit\'e Paul Sabatier, 118 Route de Narbonne,
31062 cedex, Toulouse, France}

\begin{abstract}
A complete adiabatic transport of Bose-Einstein condensate
in a double-well trap is investigated within the Landau-Zener
(LZ) and Gaussian Landau-Zener (GLZ) schemes  for the case of a
small nonlinearity, when the atomic interaction is weaker than the
coupling. The schemes use the constant (LZ) and  time-dependent Gaussian
(GLZ) couplings. The mean field calculations show that LZ and GLZ
suggest essentially different transport dynamics. Significant deviations
from the case of a strong coupling are discussed.
\end{abstract}


\maketitle

\section{Introduction}

Nowadays the trapped Bose-Einstein condensate (BEC) is one
of the most active topics in modern physics, see monograph
\cite{Pit_Str} and reviews
\cite{Dalfovo,Leggett,Yuk_lp_rew,Morsch,Bloch}.
Between many branches of this activity, investigation of
weakly bound condensates or multicomponent BEC is of keen
interest. It involves various aspects of Bose Josephson junction
\cite{Smerzi,Albiez}
including those in periodically modulated traps
\cite{holthaus_01,Weiss,Zhang_PRA_08}
and optical lattices
\cite{Morsch,Wu_Nu_00,Zobay_00},
transport problems in double-well
\cite{Weiss,Liu_PRA_02,Nest_Nov_lanl,Nest_Nov_JPB},
triple-well
\cite{Graefe_STIRAP,Wang_Chen_06,Nest_Bars_08,Nest_Nov_lanl,Nest_Nov_LP,Rab,Opatrny,Ye},
and multi-well \cite{Nistazakis_08} traps, topological states
\cite{Williams,Yuk_PRA_04}, etc. In all these areas
the nonlinearity caused by interaction between BEC atoms plays an
essential role. It drastically enriches BEC dynamics by
new effects and phenomena. At the same time,
the nonlinearity complicates and even
hampers some process, e.g. the adiabatic transport
\cite{Graefe_STIRAP, Nest_Bars_08,Nest_Nov_LP}.

In this paper, we investigate the influence of a week nonlinearity
on the adiabatic transport in a double-well trap.
The transport is produced by a controlable and irreversible
tunneling through the barrier separating the wells. It assumes that
BEC atoms, being initially in one potential well, are completely transferred
to another well and then kept there. The process can be driven
by varying the system parameters in time, such as space
separation between the wells and relative position of the well depths.
Such transport can be realized in in multi-well traps
\cite{Albiez} and arrays of selectively addressable traps \cite{arrays}.
In this problem we actually deal with two weakly bound
(through the barrier) condensates or multicomponent BEC
with the components defined as the populations of the wells.
Being produced, the complete irreversible transport could serve as a
useful tool for general manipulations of the condensate. Besides, it
could open interesting perspectives for generation and investigation
of various geometric phases \cite{Nest_Nov_LP,Balakrishnan_EPJD_06},
creation of topological states \cite{Williams,Yuk_PRA_04}, etc.

The transport in multi-well traps can be produced by
many ways: Landau-Zener
\cite{Liu_PRA_02,Wang_Chen_06,Nest_Nov_JPB}
and  Rosen-Zener  \cite{Ye} methods,
periodic time-dependent potential modulation \cite{Weiss},
Rabi switch \cite{Nistazakis_08},
Stimulated Raman Adiabatic Passage (STIRAP)
\cite{Graefe_STIRAP,Nest_Nov_lanl,Nest_Nov_LP,Rab,Opatrny}, etc..
All these methods provide a robust population transfer of the ideal
(without  interaction) condensate but
often suffer from the detrimental influence of the nonlinearity
\cite{Graefe_STIRAP,Nest_Nov_lanl,Nest_Nov_LP,Rab,Opatrny}.
This especially concerns the adiabatic population transfer methods,
like STIRAP for triple-well/level systems \cite{Berg,Vitanov,Kral},
which, being generally robust to small variations of the process
parameters, are, nevertheless, fragile to the nonlinearity.
Then a natural question arises: is it possible
to turn the nonlinearity from the detrimental to favorable
factor of the adiabatic transport?

As was recently shown \cite{Nest_Nov_JPB}, the nonlinearity can
indeed favor the adiabatic transport if it is modeled within the
Landau-Zener (LZ) \cite{Landau} and Gaussian Landau-Zener (GLZ)
\cite{Nest_Nov_JPB} protocols. The latter protocol assumes a Gaussian
time-dependent monitoring of the coupling between the wells
(barrier penetrability) $\Omega (t)$ and a linear evolution in time
of the difference $\Delta (t)$ between the well depths.
The GLZ is thus a generalization of both LZ \cite{Landau} and RZ \cite{RosZ}
methods.  It was shown \cite{Nest_Nov_JPB} that
in  LZ and GLZ the transport is asymmetric, i.e. can be considerably
enhanced or suppressed by the nonlinearity, depending on its sign (i.e.
repulsive or attractive character of the interaction).
Similar results were obtained earlier for LZ-controlled interband
transitions in accelerated optical lattices \cite{Morsch,Zobay_00}.
This useful feature of LZ-based schemes was used in \cite{Nest_Nov_JPB}
to build the successful transport protocols under a strong nonlinearity.
It was shown that, by choosing the proper monitoring, one may produce
a complete irreversible transport in wide regions of repulsive
and attractive interaction. Moreover, the transport in both left-right
and right-left directions is possible. Hence the scheme is indeed
universal. Though LZ and GLZ have much in common, the latter was shown
to be more flexible. Besides, it suggests a new transport regime where
not the center but edges of the Gaussian coupling play a decisive
role. The LZ and GLZ protocols were analyzed in terms of
nonlinear structures arising in the stationary spectra.

In this paper we continue the development of the LZ and GLZ transport
schemes but now for a case of a weak nonlinearity $UN<\Omega$ where
$N$ is the total number of BEC atoms and $U$ is the atomic interaction.
This case is not covered in \cite{Nest_Nov_JPB} and hence devotes an
additional analysis.
We will show that, at small nonlinearity,
a complete population transfer is possible for both LZ and GLZ
but, unlike the strong nonlinearity case, it takes place
at sufficiently low process rates (which was actually expected
following the previous studies
\cite{Wu_Nu_00,Zobay_00,Liu_PRA_02,Wang_Chen_06,Nest_Nov_JPB}).
At even lower rates, i.e. close to the adiabatic limit,
the LZ transfer is heavily spoiled by strong and slowly damped
Rabi oscillations while the GLZ protocol is reduced to the Rabi switch.
The nonlinear structures (loops, etc) can appear in the stationary spectra
but, unlike the strong coupling case
\cite{Nest_Nov_JPB}, they cannot be already used as
reliable tools for the analysis of transport features.
The transport asymmetry weakens for both LZ and GLZ.

This paper is outlined as follows. In Sec. II the population
transfer methods are sketched. In Sec. III the relevant mean field
formalism is done. The numerical results are
discussed in Sec. IV. The conclusions are given in the Sec. V.

\section{Transport protocols}

The LZ  tunneling  between energy levels \cite{Landau} is a general
physical process which can be straightforwardly recast for the
nonlinear transport of BEC atoms in a double-well trap
\cite{Wu_Nu_00,Zobay_00,Liu_PRA_02,Wang_Chen_06,Nest_Nov_JPB}.
The resulting scheme is illustrated in Figure 1 a)-b).
\begin{figure}[h]
\includegraphics[height=8.5cm,width=4.0cm,angle=-90]{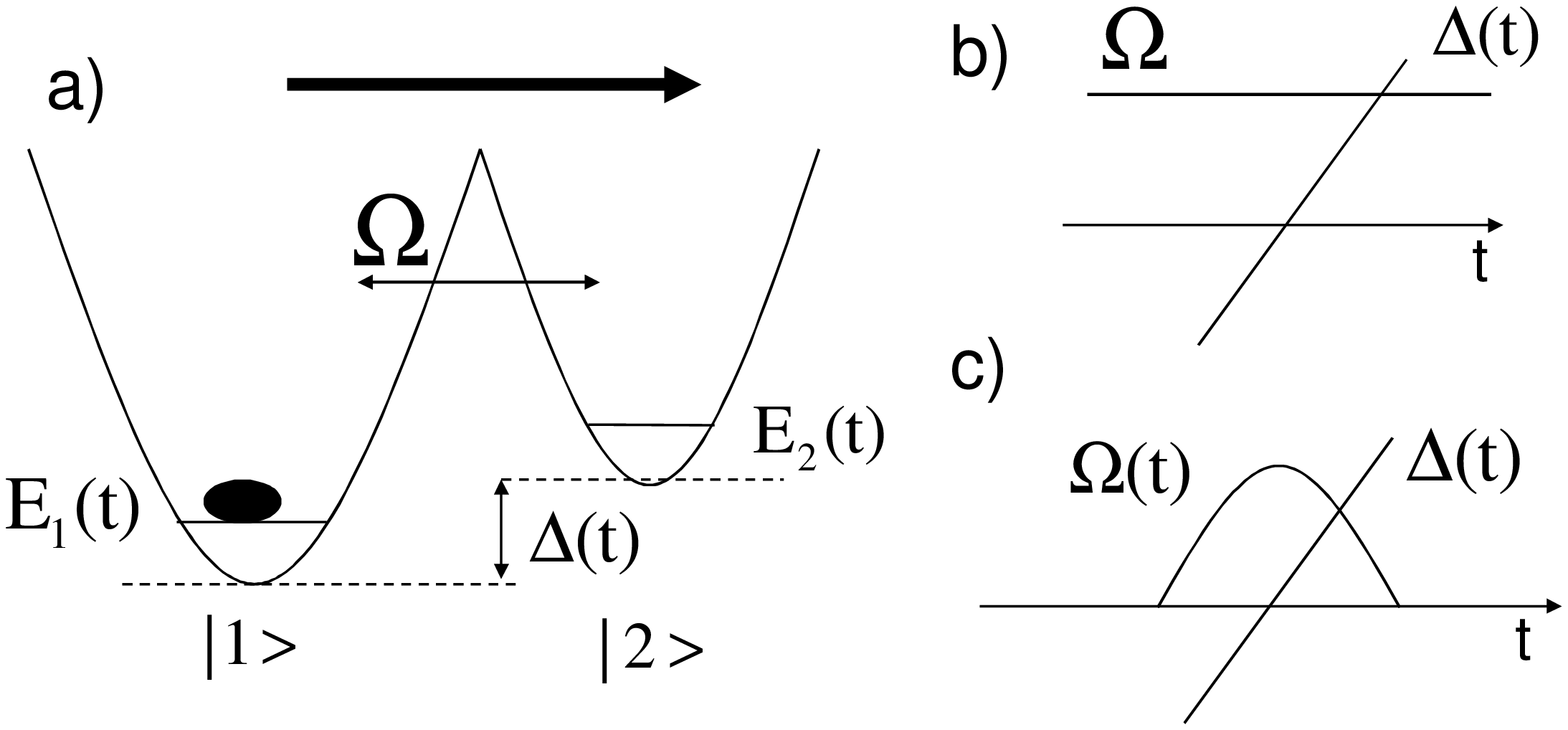}
\vspace{0.6cm} \caption{ a) The general population transfer scheme in a
double-well trap. The BEC atoms, being initially in the left well (bold dot)
are then completely transferred in the direction indicated by the upper arrow.
$E_1(t)$ and $E_2(t)$ are the ground state energies of the separated wells. The
transfer is driven by the detuning $\Delta (t)$ and coupling (barrier
penetrabulity) $\Omega $. In both LZ and GLZ, the detuning linearly depends on
time. b) The LZ protocol with the constant coupling $\Omega $. c) The GLZ
protocol with the Gaussian time-dependent coupling $\Omega (t)$. }
\end{figure}

As seen from Fig. 1a)-b), the LZ transfer is controlled by the constant
coupling $\Omega$  and time-dependent difference between the well depths
$\Delta (t)=\alpha t$ where $\alpha$ is the detuning rate. Following Fig.1c),
in the GLZ protocol, the coupling is defined as a time-dependent Gaussian pulse
\begin{equation}\label{coup}
\Omega(t)=K {\bar\Omega}(t), \quad
\bar{\Omega}(t)=\exp\{-\frac{(\bar t-t)^2}{2\Gamma^2}\},
\end{equation}
where $K$ is the amplitude, $\Gamma$ is the pulse width, and
$\bar t$ is the centroid time. In the experimental setup, the detuning can be monitored
by varying the well depths while the Gaussian coupling  by the proper
change of the separation distance between the wells \cite{arrays}.

In the linear (without nonlinearity) case, the final probability of the
LZ transfer reads \cite{Landau}
\begin{equation}\label{LZpop}
P=1-e^{-\frac{\pi\Omega^2}{2\alpha}} \; .
\end{equation}
It allows a complete transition $P=1$ only in the adiabatic
limit $a \to 0$. However, as is seen from the rough estimation
\cite{Zobay_00}
\begin{equation}\label{LZnonlin}
P\approx
1 - e^{-2\pi\frac{\Omega^2}{2\alpha}(1+\frac{\Lambda}{2\Omega})} \;,
\end{equation}
inclusion of the nonlinearity factor $\Lambda \sim U$ makes possible
the complete transport in much wider range  of $\alpha$. Moreover, the effect is
obviously asymmetric with respect to the $\Lambda$ (interaction) sign.
This peculiarity of the nonlinear LZ scheme was used in \cite{Nest_Nov_JPB}
to turn the nonlinearity from the detrimental to helpful factor for
the transport.

The introduction of the GLZ protocol is motivated by the well-known fact that
the LZ transfer actually takes place only within a finite time interval near
the symmetry point $\Delta(t)\approx 0$ when $\Delta (t) < \Omega$. Then it is
natural to use a time-dependent coupling of a certain duration, say of the
Gaussian form. Note that in fact the GLZ protocol is  a generalization of LZ
(constant coupling and time-dependent detuning) \cite{Landau} and RZ (constant
detuning and time-dependent coupling) \cite{RosZ} schemes.

\section{Mean field model}

BEC transport  is studied in mean-field approximation within the
Gross-Pitaevskii equation (GPE) \cite{Pit_eq}
\begin{equation}\label{GPE}
i\hbar{\dot \Psi}({\vec r},t) = [-\frac{\hbar^2}{2m}\nabla^2 +
V_{\rm ext}({\vec r},t) + g_0|\Psi({\vec r},t)|^2]\Psi({\vec r},t)
\end{equation}
where the dot means time derivative, $\Psi({\vec r},t)$ is the
order parameter of the system, $V_{\rm ext}({\vec r},t)$ is
the external trap potential involving both (generally
time-dependent) confinement and coupling, $g_0=4\pi a/m$ is the
parameter of the interaction between BEC atoms, $a$ is the scattering
length, and $m$ is the atomic mass.

In the two-mode approximation \cite{Milburn}, the order
parameter in a double-well trap can be written as
\begin{equation}\label{ord_par}
\Psi({\vec r},t)=\sqrt{N}(\psi_1(t)\Phi_1({\vec
r})+\psi_2(t)\Phi_2({\vec r}))
\end{equation}
where $\Phi_k({\vec r})$ is the static ground state solution of
(\ref{GPE}) for the isolated k-th well and
\begin{equation}\label{OP}
\psi_k(t)=\sqrt{N_k(t)}e^{i\phi_k(t)}
\end{equation}
is the amplitude, related to the  corresponding population
$N_k(t)$ and phase $\phi_k(t)$.
The total particle number
$N$ is conserved, i.e.
$\int d\vec r |\Psi({\vec r},t)|^2/N=\sum_{k=1}^M N_k(t) = 1$.

Note that the approximation (\ref{ord_par}) is generally valid for a weak
interaction and small number of atoms, say $N < 1000$, see discussion
\cite{holthaus_01,Milburn}. In this connection, using the GPE within the two-mode
approximation is somewhat contradictory, since the former assumes a large number
of atoms $N$ while the latter is valid for small condensates.
Nevertheless, a reasonable balance between these conditions is possible
and their combination is widely used in studies of BEC dynamics in double-well
traps, see e.g. \cite{Smerzi,Liu_PRA_02,Nistazakis_08}.

By using the linear canonical transformation \cite {Nest_Nov_LP}
\begin{equation}\label{can_trans_pop}
z=N_1-N_2, \; Z=N_1+N_2=1,
\end{equation}
\begin{equation}\label{can_trans_phase}
\theta=\frac{1}{2}(\phi_2-\phi_1), \;
\Theta=-\frac{1}{2}(\phi_1+\phi_2)
\end{equation}
it is convenient to turn the unknowns $N_k$ and $\phi_k$
to new variables, population imbalance $z$ and phase
difference $\theta$. This allow to extract from the equations
the integral of motion $Z$ and corresponding total phase $\Theta$.

Then, by substituting (\ref{ord_par})-(\ref{OP}) into (\ref{GPE}),
performing the spatial integration, and
using (\ref{can_trans_pop})-(\ref{can_trans_phase}),
we obtain  equations of motion \cite{Smerzi,Nest_Nov_LP}
\begin{equation}\label{z_final}
\dot z=-{\bar\Omega}(t)\sqrt{1-z^2}\sin{2\theta} \; ,
\end{equation}
\begin{equation}\label{phase_final}
\dot \theta=\frac{1}{2}[\Delta(t)+\Lambda z+{\bar
\Omega}(t)\frac{z}{\sqrt{1-z^2}}\cos{2\theta}] \; .
\end{equation}
where $\bar \Omega (t)= \Omega (t)/K$ is the normalized coupling (\ref{coup}),
\begin{equation}\label{detun}
\Delta (t)=\frac{1}{2K}(E_1(t)-E_2(t))=\alpha t
\end{equation}
is the scaled detuning,
\begin{equation}\label{lambda}
\Lambda=\frac{UN}{2K}
\end{equation}
is the key nonlinearity parameter determining the ratio between the coupling
amplitude $K$ and interaction $U$. In  (\ref{z_final})-(\ref{phase_final}), the
time $t$ is rescaled as $2Kt\longrightarrow t$. Eqs.
(\ref{z_final})-(\ref{phase_final}) are solved to get the numerical results
presented in the next section.

In principle, the values $\Omega(t)$, $E_{1,2}(t)$ and $U_{1,2}=U$ are
determined from the GPE as
\begin{equation}\label{Om}
 \Omega (t) = - \frac{1}{\hbar}\int d{\vec r} \;
 [\frac{\hbar^2}{2m}\nabla\Phi^*_1 \cdot\nabla\Phi_2
 +\Phi^*_2 V_{\rm ext}(t)\Phi_1] \; ,
\end{equation}
\begin{equation}\label{E}
  E_k(t)= \frac{1}{\hbar}\int d{\vec r} \;
  [\frac{\hbar^2}{2m}|\nabla\Phi^*_k|^2
  +\Phi^*_k V_{\rm ext}(t)\Phi_k] \; ,
\end{equation}
and
\begin{equation}\label{U}
  U_k= \frac{g_0}{\hbar}\int d{\vec r} \; |\Phi_k|^4 \; .
\end{equation}
However, in the present study they come as input parameters.
In GLZ, the coupling is approximated by the Gaussian function
(\ref{coup}). The energies $E_{1,2}$ enter the detuning (\ref{detun})
and the interaction $U$ between the atoms inside every well
is included into the nonlinearity parameter (\ref{lambda}).

In the previous study \cite{Nest_Nov_LP}, the stationary spectra
(chemical potentials)
\begin{equation}\label{chem_pot}
\mu=\frac{1}{2}[\Delta (t)z+\Lambda z^2
-\bar\Omega(t)\sqrt{1-z^2}\cos{2\theta}],
\end{equation}
were employed for the analysis of the transport. These spectra
are obtained by using the equations $\dot z=\dot \theta=0$
which explicitly read
\begin{equation}\label{ss_eq1}
\theta=\frac{\pi}{2}n \; ,
\end{equation}
\begin{equation}\label{ss_eq2}
\Delta (t)+z(\Lambda+(-1)^n
\frac{{\bar\Omega}(t)}{\sqrt{1-z^2}})=0
\end{equation}
where $n$ is an integer real number.
Substituting numeric solutions of (\ref{ss_eq1})-(\ref{ss_eq2})
into (\ref{chem_pot}), we get the chemical potentials
$\mu_{-}$ and $\mu_{+}$ as the eigenvalues of the
stationary states. They will be used below to demonstrate
the nonlinear structures arising in the stationary spectra.

\begin{figure*}
\includegraphics[height=12.5cm,width=11.5cm]{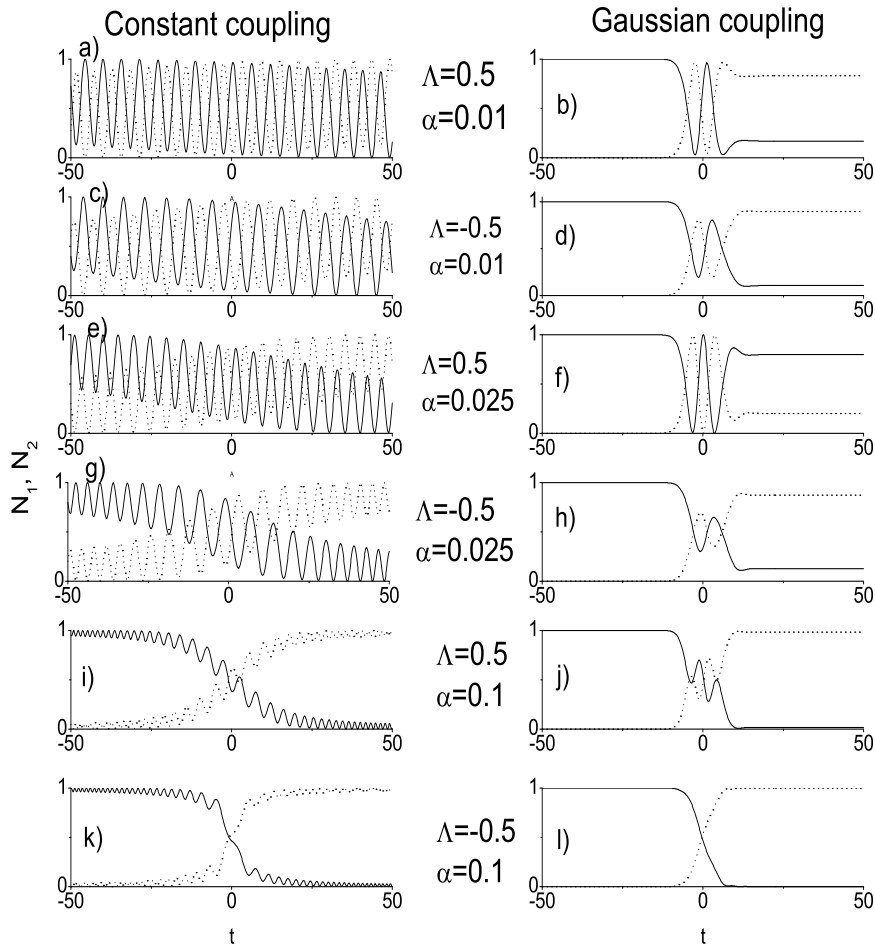}
\vspace{0.2cm}
\caption{Populations $N_1$ (dash line) and $N_2$ (solid line) at
slow detuning rates $\alpha$ and weak nonlinearity $\Lambda=\pm 0.5$
for LZ (left) and GLZ (right) protocols. Positive and negative $\Lambda$
represent repulsive and attraction interactions, respectively.}
\end{figure*}

\section{Results and discussions}

Results of our calculations are presented in the Figs. 2 and  3. In Fig. 2, the
populations of the first well, $N_1(t)$, and the second well, $N_2(t)$, are
shown for LZ and GLZ protocols. BEC atoms are initially placed in the first
well, e. g. $N_1(t\rightarrow -\infty)=1$, $N_2(t\rightarrow -\infty)=0$, and
then transferred to the second well. The weak nonlinearity $\Lambda=\pm 0.5$
covering both repulsive ($\Lambda >0$) and attractive ($\Lambda < 0$)
interaction is used. Following \cite{Zobay_00,Nest_Nov_LP}, a weak nonlinearity
allows a complete adiabatic transport only for rather small detuning rates
$\alpha$. Here we consider even smaller rates $0.01 \le \alpha \le 0.1$ to test
the vicinity of the adiabatic limit. As discussed below, just this rate region
demonstrates the major differences between LZ and GLZ protocols and is thus
most interesting for our aims. Following Fig. 2, these differences vanish while
approaching the upper value $\alpha = 0.1$, where the complete transport is
most robust, see panels i)-l). At even higher $\alpha$, the process is already
too rapid to keep the adiabaticity at the given $\Lambda$ and the transport
becomes incomplete \cite{Nest_Nov_LP}.

Figure 2 shows that the major LZ-GLZ differences take place at the lowest rate
$\alpha=0.001$ (panels a)-d)). The LZ demonstrates here high-amplitude
slowly-damped Rabi oscillations of the populations. This is because, at so
small rate, the energies $E_1(t)$ and $E_2(t)$ are almost parallel and the
condition $\Delta \le \Omega$ for Rabi oscillations is fulfilled for very long
time. The transport process in this case is vague. However, as seen from panels
e) and g), it becomes more distinctive with increasing $\alpha$.

Much better transport results take place for GLZ (panels b),d),h)) where the
Rabi oscillations exist only for a short time determined by the duration of the
Gaussian coupling $\bar\Omega (t)$. In this case, the completeness of the
transport depends on the instant when the coupling is over and Rabi
oscillations are switched off. In other words, it is determined by the width
$\Gamma$ of the Gaussian coupling pulse (\ref{coup}). Panels b), d) and d) show
that if the Rabi switch is done at the proper time, then the transport is
effective. If not (panel f)), then the transport fails. The process can be
controlled by monitoring both $\Gamma$ and $\alpha$. It is similar to other
Rabi switch techniques, e.g. \cite{Nistazakis_08}.

Panels i)-l) show that at higher rates, namely at $\alpha = 0.1$, both LZ and
GLZ demonstrate a robust and complete transport (with some advantage of the GLZ
protocol). The convergence of LZ and GLZ results is explained by the fact that
at this rate the condition $\Delta \le \Omega$ is kept in LZ already for much
shorter time comparable with the GLZ Gaussian pulse duration. This time
interval is already not enough for development of high-amplitude Rabi
oscillations. The oscillations also vanish in GLZ thus signifying the
conversion of the Rabi switch mechanism to the robust adiabatic population
transfer. As was mentioned above, a considerable further increase of $\alpha$
is not desirable since then the process will be too rapid to support the
adiabatic following.

\begin{figure*}
\includegraphics[height=13cm,width=12.5cm,angle=-90]{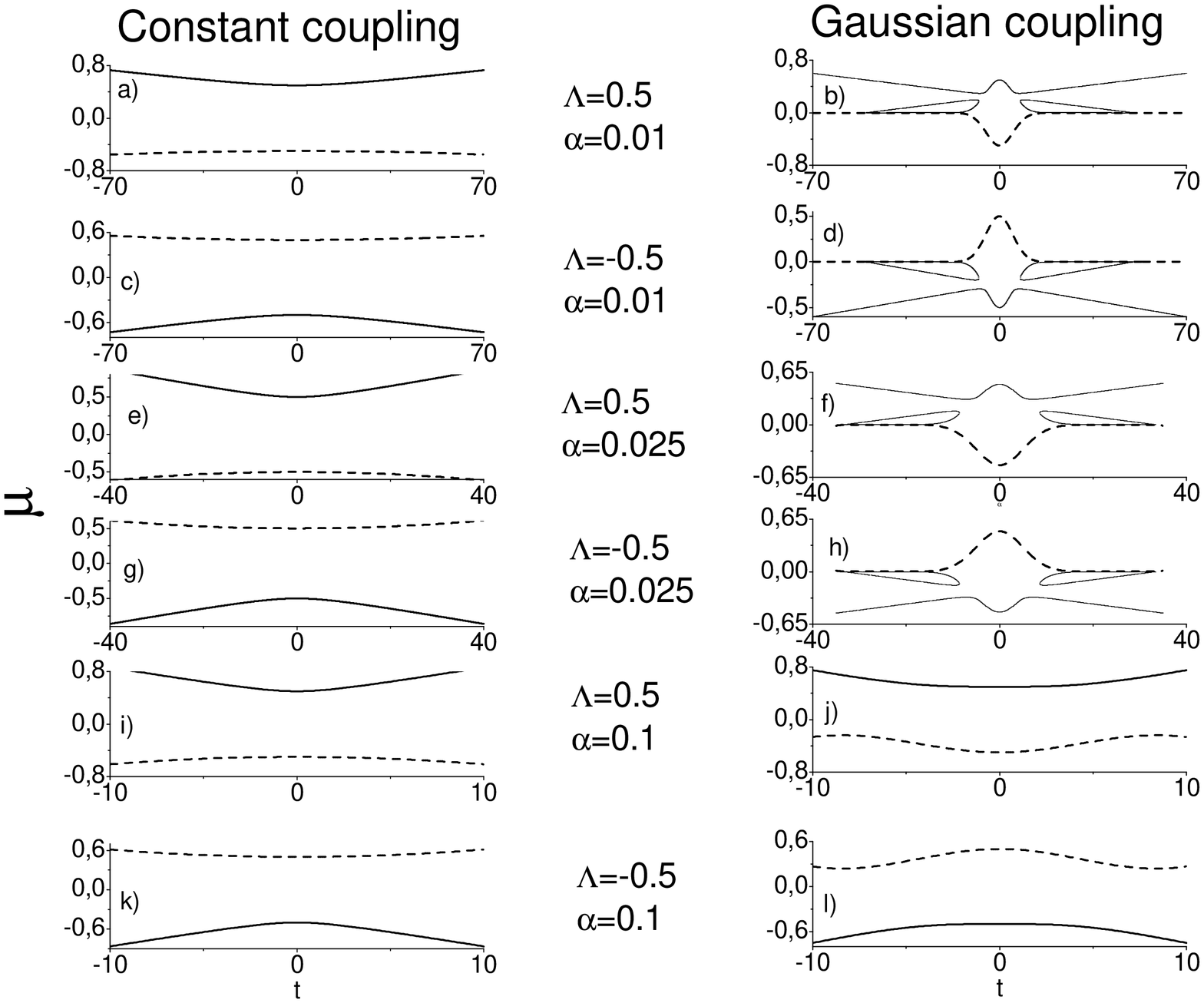}
\vspace{0.2cm} \caption{Chemical potentials $\mu_{-}$ (upper lines and
structures) and $\mu_{+}$ (lower lines and structures) for LZ (left) and GLZ
(right) protocols. The $\mu_{-}$ is depicted by solid (dash) curves for
repulsive (attractive) BEC and vise versa for $\mu_{+}$. The detuning rate
$\alpha$ and nonlinearity $\Lambda$ have the same values as in Fig. 2.}
\end{figure*}

Comparison of the results in Fig. 2 for the repulsive ($\Lambda=+0.5$) and attractive
($\Lambda=-0.5$) interaction shows that the asymmetry in BEC transport, i.e. detrimental
or favorable effect of the nonlinearity depending on the interaction sign, is generally
small (though it can manifest itself in particular fragile cases of the Rabbi switch,
see panels f) and h)). At least the effect is much weaker than for a strong nonlinearity
when it causes a drastic support or suppression of the transport \cite{Nest_Nov_LP}.
Since the asymmetry effect depends on the nonlinearity magnitude, its insignificance
for a weak nonlinearity is natural.

Altogether, Fig. 2 allows to conclude that the cases of weak and strong nonlinearity
are quite different. Though the complete adiabatic transport is possible in both cases,
a weak nonlinearity is distinguished by an essential role of the Rabi oscillations
(hence specific transport regimes like the Rabi switch) and faint asymmetry effect.

The peculiarities of weak nonlinearity are additionally demonstrated in Fig. 3
where stationary spectra $\mu_+$ and $\mu_-$ are depicted. It is seen that the
GLZ spectra exhibit  nonlinear structures (panels b),d),f),h)) despite  a weak
nonlinearity. However, these stationary spectra in general and nonlinear
structures in particular do not display the crucial role of Rabi oscillations
pertinent to a weak nonlinearity and so can hardly be used for the reliable
treatment of the transport regimes. In this sense, the LZ stationary spectra
are not instructive as well. They do not exhibit nonlinear structures at all
and so might assume a robust transport. However, as seen from Fig.2
a),c),e),d), g), the LZ transport is heavily damaged by the Rabi oscillations.
Here we see again a big difference with the case of a strong nonlinearity where
the solid nonlinear structures in the stationary spectra allow to do a reliable
analysis of the transport features.

\section{Conclusions}
The complete adiabatic transport of Bose-Einstein condensate
in a double-well trap is investigated within the conventional
Landau-Zener (LZ) protocol and its generalization to
time-dependent Gaussian coupling (GLZ) for a case of a small
nonlinearity. The relevant range of slow detuning rates when
such nonlinearity manifests its main peculiarities is analyzed
in detail. The essential role of Rabi oscillations in the vicinity
of the adiabatic limit is demonstrated. These oscillations hamper
the LZ transport but make possible the Rabi switch transport for GLZ.
For higher detuning rates, both LZ and GLZ converge to a
robust adiabatic transfer. The asymmetry effect is found generally
small.

The nonlinear structures in the stationary spectra arise at very low
detuning rates despite a small interaction.  However, these structures do
not reveal the significant
role of Rabi oscillations and so, unlike the case of a strong nonlinearity,
cannot be used as a reliable tool for the analysis of transport features.

Altogether, the calculations show a considerable difference between slightly
and strongly nonlinear transports. They demonstrate different mechanisms.
What is most important,  a weak nonlinearity with its negligible asymmetry effect
cannot be used as a powerful tool to enforce or suppress the
adiabatic transport. In this sense a strong nonlinearity is certainly more promising
\cite{Nest_Nov_LP}.

This work was supported by the grants 08-0200118 (RFBF, Russia)
and 684 (Universite Paul Sabatier, Toulouse, France, 2008).
Authors thank Dr. A.Yu. Cherny and Prof. V. I. Yukalov for useful
discussions.

\end{document}